\documentclass[%
amsmath, amssymb, aps, pra, reprint, preprintnumbers, showkeys, floatfix
]{revtex4-2}
\usepackage[utf8]{inputenc}
\usepackage[english]{babel}
\usepackage[dvips]{graphicx}
\usepackage[breaklinks=true,colorlinks,citecolor=blue,linkcolor=blue,urlcolor=blue]{hyperref}
\usepackage{xcolor,url,amsthm,amssymb,amsmath,mathtools,latexsym,appendix,physics,bbm,enumerate}
\usepackage{dcolumn}
\usepackage{bm}

\renewcommand{\exp}{\vb{e}}

\allowdisplaybreaks

\begin{document}
\preprint{SB/F/492-21}

\title[LAQC for X states]{Local available quantum correlations of X states: The symmetric and anti-symmetric cases}

\author{Hermann Albrecht}
\email[Corresponding Author: ]{albrecht@usb.ve}
\affiliation{Departamento de F\'{\i}sica, Universidad Sim\'on Bol\'{\i}var, AP 89000, Caracas 1080, Venezuela.}

\author{David Bellorín}
\affiliation{Departamento de F\'{\i}sica, Universidad Sim\'on Bol\'{\i}var, AP 89000, Caracas 1080, Venezuela.}

\author{Douglas F. Mundarain}
\affiliation{Departamento de F\'{\i}sica, Universidad Cat\'olica del Norte, Casilla 1280, Antofagasta, Chile.}

\date{April 25, 2022}

\begin{abstract}
Local available quantum correlations (LAQC), as defined by Mundarain et al., are analyzed for 2-qubit X states with local Bloch vectors of equal magnitude. Symmetric X-states are invariant under the exchange of subsystems, hence having the same {local} Bloch vector. On the other hand, anti-symmetric X states have {local} Bloch vectors with an equal magnitude but opposite direction {(anti-parallel)}. In both cases, we obtain exact analytical expressions for their LAQC quantifier. We present some examples and compare this quantum correlation to concurrence and quantum discord. We have also included Markovian decoherence, with Werner states under amplitude damping decoherence. As is the case for depolarization and phase damping, no sudden death behavior occurs for the LAQC of these states with this quantum channel.
\end{abstract}


\maketitle

\section{Introduction}

Quantum correlations and their quantification are of central importance in Quantum Information Theory (QIT). Entanglement \cite{Horodecki-Ent} was seen as the main quantum resource until the development of Quantum Discord \cite{qDiscord-Olliver, qDiscord-Henderson} in 2001. Since then, the establishment of other quantum correlations and their quantifiers \cite{Modi-qDiscord} has become a very active field of research in Information Theory, with applications in Quantum Computation and related specialties.

Local measurements play a central role in properly defining correlations. The ability of a local observer to infer the results of another one from his own must be quantified by these. Although already present in entanglement \cite{EPR-1935, schrodinger_1935, schrodinger_1936}, such a feature is at the core of defining the above-mentioned Quantum Discord:
\begin{equation}\label{eq:QD-Def}
D_A(\rho_{AB})\equiv \min_{\qty{\Pi_i^A}}\qty{I(\rho_{AB})-I\qty\big[(\Pi^A\otimes\mathbbm{1}_2)\rho_{AB}]},
\end{equation}
\noindent{}which is based on comparing the quantum Mutual Information, defined for the original state $\rho_{AB}$ as
\begin{equation}\label{eq:InfoMutua}
I\qty(\rho_{AB})\equiv S\qty(\rho_A)+S\qty(\rho_B)-S\qty(\rho_{AB}),
\end{equation}
\noindent{}where $\rho_A$ and $\rho_B$ are the reduced operators, i.e. marginals, with a corresponding post-measurement state in the absence of readout, where the measurement is performed locally on subsystem A. This latter state is referred to as a classical-quantum (or A-classical) state,
\begin{equation}\label{eq:Estados_Cl-Q}
\begin{split}
\rho^{cq}_{AB}=&\sum_i\, p_i\,\dyad{i}\otimes\rho_B^i\\ =& \sum_i\, p_i\,\Pi^{(A)}_i\otimes\rho_B^i,
\end{split}
\end{equation}
\noindent{}and all such 2-qubit states constitute a set, denoted by $\Omega_o$. An analogous set, denoted by  $\Omega_o'$, is readily defined when the local measurement is performed on the B subsystem. That is, 
\begin{equation}\label{eq:QD-Def-B}
D_B(\rho_{AB})\equiv \min_{\{\Pi_i^B\}}\left\{I(\rho_{AB})-I[(\mathbbm{1}_2\otimes\Pi^B)\rho_{AB}]\right\}\,.
\end{equation} 
Quantum correlations defined using the set $\Omega_o$ (or  $\Omega_o'$) are called Discords and are in general not symmetric so that $D_A(\rho_{AB})\neq{}D_B(\rho_{AB})$.

Another set of quantum correlations and quantifiers was developed by considering classical states, that is, post-measurement states in the absence of readout whose measurements were performed locally on both subsystems:
\begin{equation}\label{eq:Estados-Clasicos}
\rho_{AB}^c = \sum_{i,j} p_{ij} \Pi^{(A)}_i\otimes\Pi^{(B)}_j\,.
\end{equation}
\noindent{}A special subset of such states are product states,
\begin{equation}
    \rho_{AB}^{\Pi}=\sum_{i,j} p_{i} \Pi^{(A)}_i\otimes{}p_{j}\Pi^{(B)}_j=\rho^{(A)}\otimes\rho^{(B)},
\end{equation}
\noindent{}also referred to as uncorrelated states. For these, the coefficients $p_{ij}$ in eq. \eqref{eq:Estados-Clasicos} need to be factorizable, $p_{ij} = p_i p_j$. We denote the sets of classical and product 2-qubit states as $\Omega_c$ and $\Omega_{p}$, respectively.

Wu et al. considered general quantum correlations defined in terms of local bipartite measurements in \cite{Wu_ComplementaryBases} and, in a brief final appendix, outlined symmetric quantum correlations in relation to mutual information. Mundarain et al. in \cite{LAQC} developed the so-called Local Available Quantum Correlations (LAQC) by focusing on a slightly different version of those symmetric correlations, defining them in terms of mutual information of local bipartite measurements on the so-called {complementary} basis of a previously determined optimal computational basis. 

This work is focused on determining the LAQC quantifier for two particular sets of the so-called X states \cite{EstadosX}
\begin{eqnarray}\label{eq:estadosX}
\rho^X = \left(
   \begin{array}{cccc}
   \rho_{11} & 0 & 0 & \rho_{14} \\
   0 & \rho_{22} & \rho_{23} & 0 \\
   0 & \rho_{32} & \rho_{33}& 0 \\
   \rho_{41}& 0 & 0 & \rho_{44} \\
   \end{array}
  \right)\,.
\end{eqnarray}
This family of 2-qubit states has been studied extensively \cite{Quesada-XStates} and are a widely used set of 2-qubit density matrices in QIT. That is because any arbitrary 2-qubit state $\rho_{AB}$ can be mapped to such $\rho^X$, preserving its main characteristics, e.g. quantum correlations \cite{XStates-Entanglement, Hedemann-XStates}.

Experimentally, this type of 2-qubit states has been achieved with cold trapped atoms and ions \cite{XStates-ColdIons}  and in non-linear crystals \cite{XStates-NonLinearCrystal}. Also, X states appear naturally in spin chain systems when the reduced matrix of two neighboring spins is studied \cite{XStates-SpinChain, XStates-Ferromagnets} and any pure 2-qubit spin-$\frac{1}{2}$ state evolves to an X state via decoherence due to magnetic field fluctuations  \cite{Quesada-XStates}.

{An algebraic characterization of 2-qubit X states was achieved by Rau in} \cite{Rau_2009-Xstates_algebra}{, revealing an underlying symmetry of $su(2)\times{}su(2)\times{}u(1)$. Therefore, the seven operators of this $su(4)$ subalgebra are the invariance set of this family of 2-qubit states. Such invariance allows generalizing the definition of X states beyond} \eqref{eq:estadosX}{. This result facilitates characterizing quantum channels that are closed maps within this set.}

Concurrence is a \emph{bona fide} entanglement measure introduced by Wootters \cite{Wooters_Concurrence} as 
\begin{equation}\label{eq:DefConcurrencia}
\mathcal{C} \equiv \max\{0,\lambda_1-\lambda_2-\lambda_3-\lambda_4\},
\end{equation}
\noindent{}where $\qty{\lambda_i}$ are the decreasing ordered eigenvalues of $\mathbf{R}=\sqrt{\sqrt{\rho}\,\tilde{\rho}\sqrt{\rho}}$, with $\tilde{\rho} = (\sigma_y\otimes\sigma_y)\rho^*(\sigma_y\otimes\sigma_y)$, $\rho^*$ the complex conjugate of $\rho$, and $\sigma_y$ the corresponding Pauli matrix. For X states, a direct calculation shows that Concurrence takes a much simpler expression \cite{EstadosX}:
\begin{equation}\label{eq:Concurrencia_X}
\mathcal{C}_X = \frac{1}{2}\,\max \left\{0,\mathcal{C}_1,\mathcal{C}_2\right\}
\end{equation}
\noindent{where}
\begin{equation*}
\mathcal{C}_1\equiv 2\left(|\rho_{14}|-\sqrt{\rho_{22}\,\rho_{33}}\right)\qc \mathcal{C}_2\equiv 2\left(|\rho_{23}|-\sqrt{\rho_{11}\,\rho_{44}}\right).
\end{equation*}

A closed analytical expression of the quantum discord of X states has not been achieved yet. In 2010, Ali et al. \cite{Ali-QD-Xstates} proposed analytical solutions for the general 7-parameter X states, but some counterexamples were reported \cite{Chen-QD-Xstates-Ex1, Lu-QD-Xstates-Ex2, Huang-QD-Xstates-WorstCaseScenario}. Later, Vinjanampathy and Rau corrected the initial overstatement in \cite{Rau-Vinjanampathy_QD_2012}, and in 2018 Rau found expressions for higher dimensional X states \cite{Rau_QD_2018}. For a subgroup of this family of 2-qubit states, where $\rho_{22}=\rho_{33}$, some alternatives have been proposed, such as the one presented by Liao et al. \cite{Liao_QD}. In this article, we use the results of Li et al. \cite{Li-QD-Xstates}, which we briefly present in an appendix.

{We can divide the family of X states using different criteria. In particular, we can define two equivalence classes depending on whether their local Bloch vectors have equal norms. In this paper, we are focused on studying canonical X states that have equal-normed local Bloch vectors. For this equivalence class, the corresponding density matrix satisfies $\rho_{22}=\rho_{33}$ or $\rho_{11}=\rho_{44}$ in} \eqref{eq:estadosX}{. Since there are two cases for this equivalence class, we divide them into two sets, depending on whether their local Bloch vectors are parallel or anti-parallel, and call them symmetric and antisymmetric X-states, respectively.}

{This paper is structured as follows.} We start with a review of the procedure for determining the quantifier of local available quantum correlations \cite{LAQC}. Then, we study the calculation of X states in two distinct subsets: when both subsystems have the same local Bloch vector and when their local Bloch vectors have the same magnitude but opposite direction. We analyze several examples and include amplitude damping as a quantum channel for Werner states \cite{Werner}.

\section{\label{sec:LAQCs-2qubits}Local available quantum correlations of 2-qubits}

In \cite{LAQC}, the starting point for Mundarain et al. is that a density operator $\rho$ can always be written in terms of different bases. For 2-qubits, we have that
\begin{equation}\label{eq:rho_Bases}
\begin{split}
  \rho_{AB}=&\sum_{klmn}\rho_{kl}^{mn}\,\dyad{km}{ln}\\
  =&\,\sum_{ijpq}R_{ip}^{jq}\,\dyad{B(i,j)}{ B(p,q)}
  \end{split}
\end{equation}
where $k, l, m, n \in \{0,1\}$, $\qty{\ket{km}}$ is the standard computational basis, that is the basis of eigenvectors of $\sigma_z$, which is local, and $\qty{\ket{B(i,j)}}$ is another local basis that is equivalent under local unitary transformations to the former one. This equivalence can be stated as $\ket{B(i,j)}=\mathbbm{U}_A\otimes\mathbbm{U}_B\ket{ij}$, with $\mathbbm{U}_A,\mathbbm{U}_B\in\vb{U}(2)$. Any of these bases can be used as the new computational one, where we now have the basis of eigenvector of $\sigma_{\vu{u}}\equiv\va*{\sigma}\cdot\vu{u}$, with $\va*{\sigma}$ the vector whose components are the Pauli matrices and $\vu{u}\in\mathbbm{E}^3$ is a generic unitary vector. Which $\vu{u}$ to choose will depend on several conditions and/or requirements of the system at hand.

For classical states there is a local basis for which $\rho_{AB}^c$ \eqref{eq:Estados-Clasicos} is diagonal. Therefore, one can define $X_\rho\in\Omega_c$ as the classical state induced by a measurement which minimizes
\begin{equation}\label{eq:S(rho||X)}
S\qty(\rho_{AB}||X_\rho)=\min_{\Omega_c}S\qty(\rho_{AB}||\chi_\rho),
\end{equation}
\noindent{}where $S\qty(\rho||\chi)=-\mathrm{Tr}(\rho\mathrm{log}_2\chi)-S(\rho)$ is the relative entropy and
\begin{equation}\label{eq:Chi_rho}
\begin{split}
\chi_\rho =&\sum_{ij}R_{ij}\,\dyad{B(i,j)},\\ 
R_{ij} =& \expval{\rho_{AB}}{B(i,j)}\,.
\end{split}
\end{equation}
\noindent{}The minimization in \eqref{eq:S(rho||X)} is equivalent to determining the coefficients $\qty\big{R_{ij}^{opt}}$ when $\chi_\rho=X_\rho$. Such coefficients are associated with the optimal basis $\qty\big{\ket{B(i,j)}^{opt}}$ which will then serve as the new computational basis, in whose terms local available quantum correlations are then defined.

To determine the optimal basis, we start by defining the most general orthonormal basis for each subsystem:
\begin{eqnarray}\label{eq:BaseOrtonormalGen}
\ket{\mu_0^{(n)}} &=& \cos\qty(\frac{\theta_n}{2})\ket{0^{(n)}} +\sin\qty(\frac{\theta_n}{2})\exp^{i\phi_n}\ket{1^{(n)}},\\
\ket{\mu_1^{(n)}} &=& -\sin\qty(\frac{\theta_n}{2})\ket{0^{(n)}} +\cos\qty(\frac{\theta_n}{2})\exp^{i\phi_n}\ket{1^{(n)}},\nonumber
\end{eqnarray}
\noindent{}where $n=1$ denotes subsystem A and $n=2$, subsystem B. The classical correlations quantifier defined in this context in \cite{LAQC} is given by 
\begin{equation}\label{eq:CorrClasicas}
\mathcal{C}(\rho) \equiv S\left(X_\rho||\Pi_{_{X_\rho}}\right) = I(X_\rho)
\end{equation}
\noindent{}where  $\Pi_{_{X_\rho}}$ is the product state nearest to $X_\rho$ and its relative entropy can be written as the Mutual Information of state $X_\rho$, as shown by Modi et al. \cite{Modi-RelativeEntropy}. Since the mutual information \eqref{eq:InfoMutua} may be written as 
\begin{equation}\label{eq:Info_Mutua-Prob}
\begin{split}
I\qty(\rho_{AB}) =& \sum_{i,j} P_{\theta,\phi}(i_A,j_B)\\
& \qq{}\times\log_2\qty[\frac{P_{\theta,\phi}(i_A,j_B)}{P_{\theta,\phi}(i_A)P_{\theta,\phi}(j_B)}],
\end{split}
\end{equation}
\noindent{}where $P_{\theta,\phi}(i_A,j_B) = \expval{\rho_{AB}}{\mu_i^{(1)},\mu_j^{(2)}}$ are the probability distributions corresponding to $\rho_{AB}$ and $P_{\theta,\phi}(i_A) = \expval{\rho_A}{\mu_i^{(1)}}$, $P_{\theta,\phi}(j_B) = \expval{\rho_B}{\mu_i^{(2)}}$ the ones corresponding to its marginals $\rho_A$ and $\rho_B$, the required minimization of the relative entropy \eqref{eq:S(rho||X)} yields a minima for the classical correlations quantifier defined in \eqref{eq:CorrClasicas}. It is straightforward to realize that this probability distributions are directly related to the $\qty\big{R_{ij}^{opt}}$ coefficients when $\qty{\ket{\mu_i^{(1)},\mu_j^{(2)}}}$ is the optimal computational basis.

Once this new computational basis $\qty{\ket{0^{(m)}}_{opt},\ket{1^{(m)}}_{opt}}$ has been determined, the state $\rho_{AB}$ is rewritten and the complementary basis defined in terms of a new general orthonormal basis:
\begin{eqnarray}\label{eq:u0-u1}
  \ket{u_0^{(m)}}&=&\frac{1}{\sqrt{2}}\qty(\ket{0^{(m)}}_{opt}+\exp^{i\Phi_m}\ket{1^{(m)}}_{opt}),\nonumber\\
  \ket{u_1^{(m)}}&=&\frac{1}{\sqrt{2}}\qty(\ket{0^{(m)}}_{opt}-\exp^{i\Phi_m}\ket{1^{(m)}}_{opt})\,.
\end{eqnarray}
\noindent{}In this generic basis, the corresponding probability distributions, \begin{equation}\label{eq:P_phi}
  P(i_A,j_B,\Phi_1,\Phi_2) = \expval{\rho_{AB}}{u_i^{A},u_j^{B}},
\end{equation}
\noindent{}and the marginal probability distributions are determined. The maximization of $I\qty(\Phi_1,\Phi_2)$ \eqref{eq:Info_Mutua-Prob} corresponds to the LAQC quantifier:
\begin{equation}\label{eq:LAQC-quant}
    \mathcal{L}(\rho_{AB}) \equiv \max_{\qty{\Phi_1,\Phi_2}} I\qty(\Phi_1,\Phi_2)\,.
\end{equation}

\section{LAQC of X states} \label{sec:LAQCs_X}
We start by using the Bloch representation of 2-qubit states, also referred to as the Fano form or Fano-Bloch representation \cite{Fano1983}, as it allows to directly describe the so-called canonical X states in terms of five independent real parameters:
\begin{equation}\label{eq:estadosX-Bloch}
\begin{split}
\rho^X = \frac{1}{4}\Bigg( \mathbbm{1}_4 + x_3\,\sigma_3\otimes\mathbbm{1}_2& + \mathbbm{1}_2\otimes y_3\,\sigma_3\\ &+\sum_{n=1}^3 T_{n}\sigma_n\otimes\sigma_n\Bigg),
\end{split}
\end{equation}
\noindent{}where $\sigma_n$ are the Pauli matrices, $x_3=\Tr[\rho(\sigma_3\otimes\mathbbm{1}_{2})]$, $y_3=\Tr[\rho(\mathbbm{1}_{2}\otimes\sigma_3)]${, and} $T_{n}=\Tr[\rho(\sigma_n\otimes\sigma_n)]$. 

An important set of \eqref{eq:estadosX-Bloch} is comprised of those for which their local Bloch vectors are of equal magnitude $\qty(\abs{x_3}=\abs{y_3})$. Among these, Bell Diagonal states \cite{Horodecki-BD_states} constitute a special subset where $\abs{x_3}=\abs{y_3}=0$ and their LAQC were already analyzed in \cite{LAQC_BD}. For $\abs{x_3}=\abs{y_3}\neq0$, a further classification will prove useful: symmetric and anti-symmetric X states, where ${x_3}={y_3}$ and $x_3=-y_3$, respectively.

It is also worth noting that single parameter mixed states between a Bell state $\qty{\ket{\psi^\pm},\ket{\phi^\pm}}$ and an element of the computational basis $\qty{\ket{i,j}}$ belong to this set of X states. These can be thought of as analogous to what Werner states \cite{Werner} are to Bell Diagonal ones. That is, as a one-parameter subset of the latter.

\subsection{Symmetric X States}

The set of symmetric X states $\rho^{X_s}$ is invariant under subsystem exchange $\vb{A}\leftrightarrow\vb{B}$ since their local Bloch parameters are equal,  ${x_3}={y_3}$. Such states arise when studying amplitude decoherence of Werner states. They also turn out as the ground state within the approximation of nanoelectric \textit{LC}-circuits as two coupled harmonic oscillators presented in \cite{Fedorov_2015}.

We start our analysis by determining the previously defined coefficients $R_{ij}(\theta_1,\theta_2,\phi_1,\phi_2)$ \eqref{eq:Chi_rho}, obtaining
\begin{eqnarray}\label{eq:Rij-XS-12}
        R_{00}&=& \,\frac{1}{4} +\frac{1}{4}\qty(\cos\theta_1+\cos\theta_2)x_3\nonumber\\ &&+\,\vb*{\Lambda}\sum_{m=0}^{1}\cos\qty[\phi_1+(-1)^m\phi_2]\qty[T_1-(-1)^mT_2]\nonumber\\
        &&+\frac{1}{4}\cos\theta_1\cos\theta_2\,T_3,\nonumber\\
        R_{01}&=&\,\frac{1}{4} +\frac{1}{4}\qty(\cos\theta_1-\cos\theta_2)x_3\nonumber\\ &&-\,\vb*{\Lambda}\sum_{m=0}^{1}\cos\qty[\phi_1+(-1)^m\phi_2]\qty[T_1-(-1)^mT_2]\nonumber\\
        &&-\frac{1}{4}\cos\theta_1\cos\theta_2\,T_3,\\
        R_{10}&=&\,\frac{1}{4} -\frac{1}{4}\qty(\cos\theta_1-\cos\theta_2)x_3\nonumber\\ &&-\,\vb*{\Lambda}\sum_{m=0}^{1}\cos\qty[\phi_1+(-1)^m\phi_2]\qty[T_1-(-1)^mT_2]\nonumber\\
        &&-\frac{1}{4}\cos\theta_1\cos\theta_2\,T_3,\nonumber\\
        R_{11}&=&\,\frac{1}{4} -\frac{1}{4}\qty(\cos\theta_1+\cos\theta_2)x_3\nonumber\\ &&+\,\vb*{\Lambda}\sum_{m=0}^{1}\cos\qty[\phi_1+(-1)^m\phi_2]\qty[T_1-(-1)^mT_2]\nonumber\\
        &&+\frac{1}{4}\cos\theta_1\cos\theta_2\,T_3\,.\nonumber
\end{eqnarray}
\noindent{}where 
\begin{equation}\label{eq:Lambda}
    \vb*{\Lambda}\equiv \frac{1}{2}\prod_{i=1}^{2}\sin\qty(\frac{\theta_i}{2})\cos\qty(\frac{\theta_i}{2})\,.
\end{equation}

Since symmetric X states are invariant under subsystem exchange $\vb{A}\leftrightarrow\vb{B}$, the angles $\theta_i$ and $\phi_i$ must maintain this symmetry in \eqref{eq:Rij-XS-12}. Therefore, only two angles are sufficient to determine the optimal computational basis. By defining $\theta_1=\theta_2=\theta$ and $\phi_1=\phi_2=\phi$ as the respective common angles, the previous expressions are readily simplified:
\begin{eqnarray}\label{eq:Rij-XS}
    R_{00}(\theta,\phi)&=& \frac{1}{4}\big[1 +2\cos\theta\,x_3+\sin^2\theta\big(\cos^2\phi\,T_1\nonumber\\ &&\quad+\sin^2\phi\,T_2\big) +\cos^2\theta\,T_3\big],\nonumber\\
    R_{01}(\theta,\phi)&=&\frac{1}{4}\big[1 - \sin^2\theta\qty(\cos^2\phi\,T_1+\sin^2\phi\,T_2)\nonumber\\ &&-\cos^2\theta\,T_3\big] =\,R_{10}(\theta,\phi),\\
    R_{11}(\theta,\phi)&=&\frac{1}{4}\big[1 -2\cos\theta\,x_3+\sin^2\theta\big(\cos^2\phi\,T_1\nonumber\\ &&\quad+\sin^2\phi\,T_2\big) +\cos^2\theta\,T_3\big],\nonumber
\end{eqnarray}
The corresponding coefficients $\qty{R_{i_A}(\theta,\phi),R_{j_B}(\theta,\phi)}$ for the reduced matrices are given by
\begin{equation}
\begin{split}
    R_{0_A}(\theta)&=\frac{1}{2}\qty(1+x_3\cos\theta) =R_{0_B}(\theta),\\
    R_{1_A}(\theta)&=\frac{1}{2}\qty(1-x_3\cos\theta)=R_{1_B}(\theta).
\end{split}
\end{equation}

The necessary minimization of the classical correlations quantifier \eqref{eq:CorrClasicas} leads to three cases to be considered:
\begin{equation}\label{eq:Cases-Angles-Xs}
    \begin{cases}
    &\theta=0\qc\phi\in\qty[0,2\pi],\\
    &\theta=\frac{\pi}{2}\qc\phi=0,\\
    &\theta=\frac{\pi}{2}\qc\phi=\frac{\pi}{2}.
    \end{cases}
\end{equation}
After defining the following functions
\begin{subequations}\label{eq:g1g2g+}
\begin{align}
   g_1\equiv\,&  \frac{1+T_1}{2}\log_2\qty(1+T_1) + \frac{1-T_1}{2}\log_2\qty(1-T_1),\label{eq:g1(T1)}\\
   g_2\equiv\,& \frac{1+T_2}{2}\log_2\qty(1+T_2) + \frac{1-T_2}{2}\log_2\qty(1-T_2),\label{eq:g2(T2)}\\
   g_{+}\equiv\,& \frac{1+T_3 + 2 x_3}{4}\log_2\qty[\frac{1+T_3 + 2 x_3}{(1+x_3)^2}]\nonumber\\
    &\;\;+\frac{1+T_3 - 2 x_3}{4}\log_2\qty[\frac{1+T_3 - 2 x_3}{(1-x_3)^2}]\nonumber\\ 
    &\;\;+ \frac{1-T_3}{2}\log_2\qty(\frac{1-T_3}{1-x_3^2})\label{eq:g+},
\end{align}
\end{subequations}
the classical correlations' quantifier \eqref{eq:CorrClasicas} can be written as
\begin{equation}\label{eq:CorrClas-XS}
 \mathcal{C}\qty(\rho^{X_s}) = \min\qty(g_1,g_2,g_{+}),
\end{equation}
so that
\begin{equation}\label{eq:OptimalBasis-Angles-Xs}
    \begin{cases}
    &\mathcal{C}\qty(\rho^{X_s}) = g_{+}\;\Longrightarrow\; \theta=0\qc\phi\in\qty[0,2\pi],\\
    &\mathcal{C}\qty(\rho^{X_s}) = g_{1}\;\Longrightarrow\;\theta=\frac{\pi}{2}\qc\phi=0,\\
    &\mathcal{C}\qty(\rho^{X_s}) = g_{2}\;\Longrightarrow\;\theta=\frac{\pi}{2}\qc\phi=\frac{\pi}{2}.
    \end{cases}
\end{equation}

After the $\theta$ and $\phi$ angles have been determined with \eqref{eq:OptimalBasis-Angles-Xs}, the state $\rho^{X_s}$ is rewritten in the optimal computational basis as to determine the probability distributions $P^{\qty(\theta,\phi)}\qty(i_A,j_B,\Phi)$ \eqref{eq:P_phi} as well as the marginal distributions $P_A^{\qty(\theta,\phi)}\qty(i_A,\Phi)$ and $P_B^{\qty(\theta,\phi)}\qty(i_B,\Phi)$. For each of the above cases, the following results are obtained:

\begin{itemize}
    \item $\theta=0,\phi\in[0,2\pi]$: Since $\phi$ always appears as $\Psi\equiv\Phi-\phi$ $\qty(\Psi\in[-2\pi,2\pi])$, we can use this new angle as the maximization parameter. For simplicity and without loss of generality, we fix $\phi=0$, so that
    \begin{equation}\label{eq:Pij-XS-00}
   \begin{split}
    P^{(0,0)}\qty(0,0,\Phi)=& \frac{1}{4}\qty(1+T_1\cos^2\Phi+T_2\sin^2\Phi),\\ =&\, P^{(0,0)}\qty(1,1,\Phi),\\
    P^{(0,0)}\qty(1,0,\Phi)=& \frac{1}{4}\qty(1-T_1\cos^2\Phi-T_2\sin^2\Phi),\\ =&\, P^{(0,0)}\qty(0,1,\Phi),
  \end{split}
    \end{equation}
    and 
    \begin{equation}\label{eq:Pj-XS-00}
    \begin{split}
        P_A^{(0,0)}\qty(0)&= P_A^{(0,0)}\qty(1) = \frac{1}{2},\\ P_B^{(0,0)}\qty(0) &= P_B^{(0,0)}\qty(1)= \frac{1}{2}\,.
    \end{split}
    \end{equation}
    \item $\theta=\frac{\pi}{2},\phi=0$:
    \begin{equation}\label{eq:Pij-XS-Pi20}
  \begin{split}
       P^{\qty(\frac{\pi}{2},0)}\qty(0,0,\Phi)=& \frac{1}{4} \big(1 + T_2\sin^2\Phi+T_3\cos^2\Phi\\
       & \qq{}-2x_3\cos\Phi\big),\\
       P^{\qty(\frac{\pi}{2},0)}\qty(1,0,\Phi)=& \frac{1}{4} \qty(1 - T_2\sin^2\Phi -T_3\cos^2\Phi),\\
    =&\, P^{\qty(\frac{\pi}{2},0)}\qty(0,1,\Phi) ,\\
       P^{\qty(\frac{\pi}{2},0)}\qty(1,1,\Phi)=& \frac{1}{4} \big(1 + T_2\sin^2\Phi+T_3\cos^2\Phi\\
       & \qq{}+2x_3\cos\Phi\big)\,.
  \end{split}
    \end{equation}
    and 
    \begin{equation}\label{eq:Pj-XS-Pi0}
    \begin{split}
        P_A^{\qty(\frac{\pi}{2},0)}\qty(0)&= P_B^{\qty(\frac{\pi}{2},0)}\qty(0) = \frac{1}{2}\qty(1 -x_3\cos\Phi),\\ P_A^{\qty(\frac{\pi}{2},0)}\qty(1) &= P_B^{\qty(\frac{\pi}{2},0)}\qty(1)= \frac{1}{2}\qty(1+x_3\cos\Phi).
    \end{split}
    \end{equation}
    \item $\theta=\frac{\pi}{2},\phi=\frac{\pi}{2}$:
    \begin{equation}\label{eq:Pij-XS-Pi2Pi2}
  \begin{split}
       P^{\qty(\frac{\pi}{2},\frac{\pi}{2})}\qty(0,0,\Phi)=& \frac{1}{4} \big(1 + T_1\sin^2\Phi+T_3\cos^2\Phi\\
       & \qq{}-2x_3\cos\Phi\big),\\
       P^{\qty(\frac{\pi}{2},\frac{\pi}{2})}\qty(1,0,\Phi)=& \frac{1}{4} \qty(1 - T_1\sin^2\Phi -T_3\cos^2\Phi),\\
    =&\, P^{\qty(\frac{\pi}{2},\frac{\pi}{2})}\qty(0,1,\Phi) ,\\
       P^{\qty(\frac{\pi}{2},\frac{\pi}{2})}\qty(1,1,\Phi)=& \frac{1}{4} \big(1 + T_1\sin^2\Phi+T_3\cos^2\Phi\\
       & \qq{}+2x_3\cos\Phi\big)\,.
  \end{split}
  \end{equation}
    and again we have that
    \begin{equation}\label{eq:Pj-XS-PiPi}
    \begin{split}
        P_A^{\qty(\frac{\pi}{2},\frac{\pi}{2})}\qty(0)&= P_B^{\qty(\frac{\pi}{2},\frac{\pi}{2})}\qty(0) = \frac{1}{2}\qty(1 -x_3\cos\Phi),\\ P_A^{\qty(\frac{\pi}{2},\frac{\pi}{2})}\qty(1) &= P_B^{\qty(\frac{\pi}{2},\frac{\pi}{2})}\qty(1)= \frac{1}{2}\qty(1+x_3\cos\Phi).
    \end{split}
    \end{equation}
\end{itemize}

To determine the LAQC quantifier \eqref{eq:LAQC-quant}, the above expressions need to be maximized for $\Phi\in\qty[0,2\pi]$. For $\theta=0,\phi=0$, we have that the probability distributions $P^{(0,\phi))}(i,j)$ depend only on $T_1$ and $T_2$, with the marginal distributions $P^{(0,\phi))}_{A,B}(i)$ equal to one-half. For $\theta=\frac{\pi}{2},\phi=0$, the Bloch parameters involved are $T_2$, $T_3$, and $x_3$, and finally, when $\theta=\frac{\pi}{2},\phi=\frac{\pi}{2}$ the only Bloch parameter that is not involved in the probability distributions and marginals is $T_2$. Therefore, we have that
\begin{equation}\label{eq:LAQC-EstadosX-s}
    \mathcal{L}\qty(\rho^{X_s}) = \max\qty(g_1,g_2,g_{+}).
\end{equation}

\subsubsection{Werner states under amplitude damping decoherence}\label{sec:Werner_AD}

Werner states, $\rho_w$, can be written as:
\begin{equation}\label{eq:rhoWerner}
\rho_w= z\dyad{\Psi^-} +\frac{1-z}{4}\,\mathbbm{1}_4,
\end{equation}
\noindent{}where $z\in[0,1]$ and $\ket{\Psi^-}=\frac{1}{\sqrt{2}}\qty(\ket{01}-\ket{10})$ is the singlet, one of the four maximally entangled 2-qubit states known as Bell states. 

The amplitude damping quantum channel \cite{Nielsen-QIT} describes the process of energy dissipation into the environment, like the $T_1$ process in NMR. For this reason it is of great importance in quantum information theory and its applications. In a previous article \cite{LAQC_BD, LAQC_BD-Err}, we analyzed the behavior of LAQC for such states under depolarizing and phase damping channels, assuming the same channel parameter for each subsystem. We did not include amplitude damping since the resulting state no longer belongs to the Bell Diagonal set but rather to the symmetric X states.

Within the Kraus operators formalism \cite{Kraus-Article}, the effect of the environment for a 2-qubit state is described via
\begin{equation}\label{eq:Kraus_Interaccion}
  \rho\longrightarrow \rho'=\sum_{i,j} \left(\vb{E}^{(A)}_i\otimes\vb{E}^{(B)}_j\right)\rho\left(\vb{E}^{(A)}_i\otimes\vb{E}^{(B)}_j\right)^\dagger,
\end{equation}
\noindent{}where the type of interaction for each subsystem may not be the same or even the  parameter describing a common type of interaction might be different. In our present analysis, we study an equal interaction parameter for both subsystems.

The Kraus operators for this quantum channel are
\begin{equation}\label{eq:Kraus_AD}
    \vb{E}^{(AD)}_0 = \mqty(1 & 0\\ 0 & \sqrt{1-p})\,\qc\quad \vb{E}^{(AD)}_1 = \mqty(0 & \sqrt{p}\\ 0 & 0),
\end{equation}
\noindent{}where $p$, the channel parameter, can be thought of as the probability of loosing a single quantum of energy, i.e. $\ket{1} \rightarrow \ket{0}$.

The state $\rho_w^{(AD)}$ resulting from applying the above Kraus operators via \eqref{eq:Kraus_Interaccion} is a symmetric X state and has the following non-null Bloch parameters:
\begin{equation}\label{eq:WernerAD-Bloch}
\begin{split}
    x_3=y_3=p\,\qc &T_1=T_2=-(1-p)z,\\ 
    &T_3=p^2 -(1-p)^2z\,.
\end{split}
\end{equation}

To determine, our correlations quantifiers \eqref{eq:CorrClas-XS} and \eqref{eq:LAQC-EstadosX-s}, we substitute these parameters in our expressions \eqref{eq:g1g2g+}. Since $T_1=T_2$, we also have that $g_1=g_2$ so that
\begin{subequations}
\begin{align}
g_{+}^{w_{AD}}(z,p)=& \frac{1}{4}\qty(1-p)^2(1-z)\log_2(1-z)\nonumber\\
    &+\frac{1}{4}\qty[\qty(1-z)(1+p^2)+2(1+z)p]\nonumber\\
    &\qq{}\times\log_2\qty[\qty(1-z)(1+p^2)+2(1+z)p]\nonumber\\ 
    &+\frac{1}{2}(1-p)\qty[(1-z)p+z+1]\nonumber\\
    &\qq{}\times\log_2\qty[(1-z)p+z+1]\nonumber\\
    &-(p+1)\log_2(1+p)\label{eq:WAD-g+},\\
g_{1}^{w_{AD}}(z,p)=&\frac{1}{2}\qty[1+(1-p)z]\log_2\qty[1+(1-p)z]\label{eq:WAD-g1}\\
&+\frac{1}{2}\qty[1-(1-p)z]\log_2\qty[1-(1-p)z].\nonumber
\end{align}
\end{subequations}

\begin{figure}[ht]
\centering
\includegraphics[width=.35\textwidth]{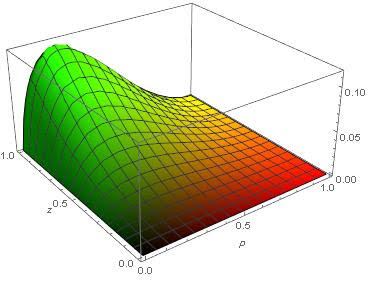}
\caption{Surface $\mathcal{S}(z,p)$ \eqref{eq:S'(z,p)} showing the difference of the $g_{1}^{w_{AD}}(z,p)$ and $g_{+}^{w_{AD}}(z,p)$ for a Werner state under amplitude damping.\label{fig:g1g+-WernerAD}}
\end{figure}

In Figure \ref{fig:g1g+-WernerAD} we present the surface $\mathcal{S}(z,p)$ that results from the difference of $g_{1}^{w_{AD}}(z,p)$ \eqref{eq:WAD-g1} and $g_{+}^{w_{AD}}(z,p)$ \eqref{eq:WAD-g+}. That is,
\begin{equation}\label{eq:S(z,p)}
    \mathcal{S}(z,p) = g_{1}^{w_{AD}}(z,p) - g_{+}^{w_{AD}}(z,p).
\end{equation}

Since $\mathcal{S}(z,p)\geq0$, we can immediately conclude that 
\begin{equation}\label{eq:C&L-WAD}
    \mathcal{C}\qty(\rho_w^{(AD)}) = g_{+}^{w_{(AD)}}\qc \mathcal{L}\qty(\rho_w^{(AD)}) = g_{1}^{w_{AD}}.
\end{equation}
A simple calculation shows that $\mathcal{L}\qty(\rho_w^{(AD)})$ is only zero when $p=0\,\forall\,z$ and $z=0\,\forall\,p$.

The Concurrence $\mathcal{C}_{w^{AD}}$ \eqref{eq:Concurrencia_X} for an X state with Bloch parameters \eqref{eq:WernerAD-Bloch} is given by
\begin{equation}\label{eq:Concurrencia_WernerAD}
\begin{split}
    \mathcal{C}_{w^{AD}}=&\max\qty\big[0,\,\mathcal{C}_2],\qq{with}\\  \mathcal{C}_2=&\frac{1-p}{2}\Bigg[2z -\sqrt{1-z}\sqrt{\qty(1+p)^2-\qty(1-p)^2z}\Bigg],
\end{split}
\end{equation}
while its Quantum Discord is given by \eqref{eq:QD-Examples} using the corresponding Bloch parameters \eqref{eq:WernerAD-Bloch}.

\begin{figure}[t]
\centering
\includegraphics[width=0.3\textwidth]{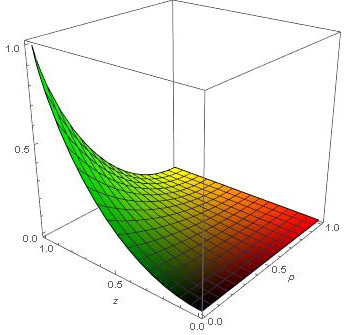}\\
\includegraphics[width=0.3\textwidth]{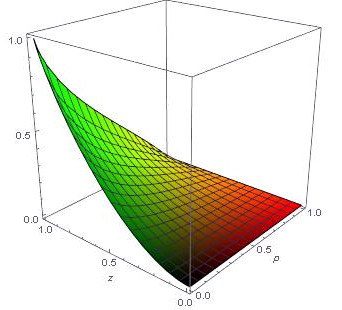}\\
\includegraphics[width=0.3\textwidth]{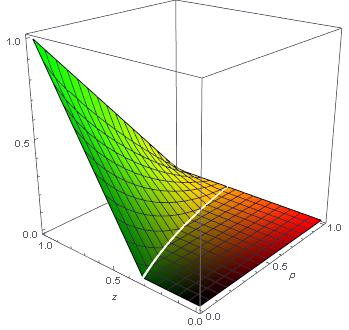}
\caption{LAQC (above), QD (middle), and Concurrence (below) for a Werner state under amplitude damping decoherence.\label{fig:Werner-ADChannel}}
\end{figure}

\begin{figure}[b]
\centering
\includegraphics[width=0.35\textwidth]{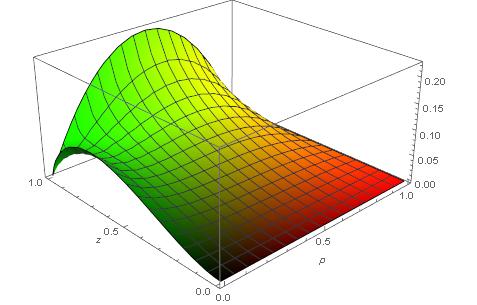}
\caption{Surface $\mathcal{S}'(z,p)$ \eqref{eq:S'(z,p)} resulting from the difference between QD and LAQC for a Werner state under amplitude damping.\label{fig:Werner-ADChannel-QD-LAQC}}
\end{figure}

In Figure \ref{fig:Werner-ADChannel} we present the graphical behavior of LAQC, QD, and concurrence. As expected, Werner states exhibit the so called `Entanglement Sudden Death' \cite{EntSuddenDeath}, while LAQC only vanishes asymptotically, as does the QD. That was already observed for two other decoherent interactions (depolarization and phase damping) in \cite{LAQC_BD}, therefore LAQC appears to be more robust as a quantum resource than entanglement, at least for this family of BD states. In Figure \ref{fig:Werner-ADChannel-QD-LAQC} we can observe the difference between QD and LAQC for this type of states. That is, we show the surface 
\begin{equation}\label{eq:S'(z,p)}
    \mathcal{S}'(z,p)=D_A\qty(\rho_w^{(AD)}) -\mathcal{L}\qty(\rho_w^{(AD)}).
\end{equation}

\subsubsection{A single-parameter symmetric X state}

As was previously stated, using Bell states and basis vectors, we can define a single-parameter symmetric X state. As an example, we have the following density operator
\begin{equation}\label{eq:Ejemplo-X_S}
 \rho_s = F \dyad{\Psi^-} +(1-F)\dyad{00},
\end{equation}
whose non-zero Bloch parameters are given by
\begin{equation}\label{eq:Bloch-Ejemplo-X_S}
\begin{split}
    x_3 = y_3= 1-F\,\qc &T_1=T_2=-F,\\ 
    &T_3=1-2F\,.    
\end{split}
\end{equation}

For this state the expressions \eqref{eq:g1g2g+} are given by
\begin{subequations}
\begin{align}
    g_{+}^{s}(F)=&\, 2-F -F\log_2\qty(2-F)\nonumber\\
    &\;+(1-F)\log_2\qty[\frac{1-F}{(2-F)^2}],\label{eq:g+-EjXS}\\
    g_{1}^{s}(F)=&\, \frac{1}{2}\big[(1-F)\log_2(1-F)\nonumber\\
    &\qq{} +(1+F)\log_2(1+F)\big]\label{eq:g1-EjXS}
\end{align}
\end{subequations}

\begin{figure}[ht]
\centering
\includegraphics[width=.35\textwidth]{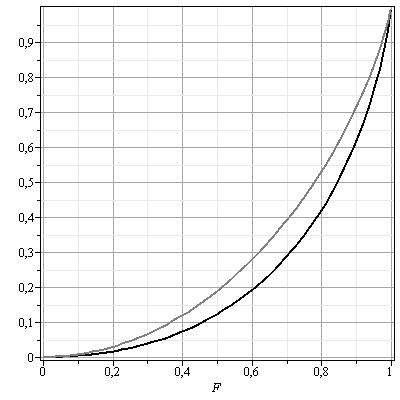}
\caption{Graphical comparison of the functions $g_{+}^{s}(F)$ (black) and $g_{1}^{s}(F)$ (gray).\label{fig:CorrClas-EjXS}}
\end{figure}

It is easily verified that $g_{1}^{s}\geq{}g_{+}^{s}$ (see Figure \ref{fig:CorrClas-EjXS}). Therefore, the correlation quantifiers for state \eqref{eq:Ejemplo-X_S} are given by
\begin{equation}\label{eq:C&L-Ejemplo-X_S}
    \mathcal{C}\qty(\rho_s) = g_{+}^{s}(F)\qc \mathcal{L}\qty(\rho_s) = g_{1}^{s}(F).
\end{equation}

The Concurrence for \eqref{eq:Ejemplo-X_S} is given by
\begin{equation}\label{eq:Concurrencia_Ejemplo-X_S}
    \mathcal{C}_s=F,
\end{equation}
while its Quantum Discord, computed using the results presented in \cite{Li-QD-Xstates} and discussed in the appendix, is given by \eqref{eq:QD-Examples} using the corresponding Bloch parameters \eqref{eq:Bloch-Ejemplo-X_S}.

\begin{figure}[b]
    \centering
    \includegraphics[width=0.4\textwidth]{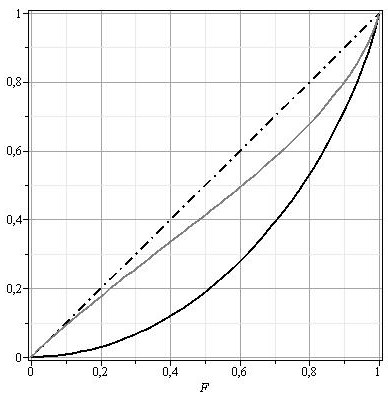}
    \caption{LAQC (continuous black line), QD (continuous gray line), and Concurrence (dash-dot line) for state \eqref{eq:Ejemplo-X_S}.\label{fig:Ejemplo-X_S}}
\end{figure}

In Figure \ref{fig:Ejemplo-X_S} the above-mentioned quantifiers are presented. As can be observed, $\mathcal{L}\qty(\rho_s)<D_A(\rho_s)<\mathcal{C}_s$ for $F\in\qty(0,1)$ and are only equal at $F=0$ and $F=1$. 

In \cite{Ali-QD-Xstates}, the authors present an example with a symmetric X state similar to \eqref{eq:Ejemplo-X_S}, namely
\begin{equation}
    \rho_{ARA}=F \dyad{\Psi^+} +(1-F)\dyad{11},
\end{equation}
whose non-null Bloch parameters are given by
\begin{equation}
\begin{split}
    x_3 = y_3= F-1\,\qc &T_1=T_2=F,\\ 
    &T_3=1-2F.    
\end{split}
\end{equation}
For such a state, the concurrence is also given by \eqref{eq:Concurrencia_Ejemplo-X_S}. Ali et al. present their results for the QD, Concurrence, and their classical correlations quantifier in a graph (see Fig. 1 in \cite{Ali-QD-Xstates}) similar to Figure \ref{fig:Ejemplo-X_S}. This similarity also occurs for Werner states (see Fig. 1 in \cite{LAQC_BD} and Fig. 3 in \cite{Ali-QD-Xstates}). Although their result for the classical correlations quantifier and the LAQC quantifier have similar graphical behavior in both cases, it is worth noticing that they measure different types of correlations.

\subsection{Anti-symmetric X States}

We now focus on studying anti-symmetric X states, i.e. $\rho^X$ whose local Bloch vectors have equal norm but opposing direction. In the Fano-Bloch representation \eqref{eq:estadosX-Bloch}, for such states we have that $x_3=-y_3$. The $R_{ij}(\theta_1,\theta_2,\phi_1,\phi_2)$ coefficients \eqref{eq:Chi_rho} previously defined are readily determined:

\begin{eqnarray}\label{eq:Rij-XAS-12}
        R_{00}&=& \,\frac{1}{4} +\frac{1}{4}\qty(\cos\theta_1-\cos\theta_2)x_3\nonumber\\ &&+\,\vb*{\Lambda}\sum_{m=0}^{1}\cos\qty[\phi_1+(-1)^m\phi_2]\qty[T_1-(-1)^mT_2]\nonumber\\
        &&+\frac{1}{4}\cos\theta_1\cos\theta_2\,T_3,\nonumber\\
        R_{01}&=&\,\frac{1}{4} +\frac{1}{4}\qty(\cos\theta_1+\cos\theta_2)x_3\nonumber\\ &&-\,\vb*{\Lambda}\sum_{m=0}^{1}\cos\qty[\phi_1+(-1)^m\phi_2]\qty[T_1-(-1)^mT_2]\nonumber\\
        &&-\frac{1}{4}\cos\theta_1\cos\theta_2\,T_3,\\
        R_{10}&=&\,\frac{1}{4} -\frac{1}{4}\qty(\cos\theta_1+\cos\theta_2)x_3\nonumber\\ &&-\,\vb*{\Lambda}\sum_{m=0}^{1}\cos\qty[\phi_1+(-1)^m\phi_2]\qty[T_1-(-1)^mT_2]\nonumber\\
        &&-\frac{1}{4}\cos\theta_1\cos\theta_2\,T_3,\nonumber\\
        R_{11}&=&\,\frac{1}{4} -\frac{1}{4}\qty(\cos\theta_1-\cos\theta_2)x_3\nonumber\\ &&+\,\vb*{\Lambda}\sum_{m=0}^{1}\cos\qty[\phi_1+(-1)^m\phi_2]\qty[T_1-(-1)^mT_2]\nonumber\\
        &&+\frac{1}{4}\cos\theta_1\cos\theta_2\,T_3\,.\nonumber
\end{eqnarray}
\noindent{}where $\vb*{\Lambda}$ is as defined in \eqref{eq:Lambda}.

As with the symmetric case, a relation between the angles $\theta_i$ and $\phi_i$ is expected. A direct calculation exchanging local angles $\theta_i$ and $\phi_i$ leads to
\begin{equation}
\begin{split}
    R_{ii}(\theta_1,\theta_2,\phi_1,\phi_2) &= R_{ii}(\theta_2+\pi,\theta_1+\pi,\phi_1,\phi_2),\\
    R_{ij}(\theta_1,\theta_2,\phi_1,\phi_2) &= R_{ij}(\theta_2,\theta_1,\phi_1,\phi_2)\qc{}i\neq{j}.
\end{split}
\end{equation}
Therefore, as for the symmetric case, we define $\theta$ and $\phi$ as the respective common local angles and the expressions in \eqref{eq:Rij-XAS-12} are readily simplified:
\begin{eqnarray}\label{eq:Rij-XAS}
    R_{00}(\theta,\phi)&=&\frac{1}{4}\big[1 + \sin^2\theta\qty(\cos^2\phi\,T_1+\sin^2\phi\,T_2)\nonumber\\ &&+\cos^2\theta\,T_3\big] =\,R_{11}(\theta,\phi),\nonumber\\
    R_{01}(\theta,\phi)&=&\frac{1}{4}\big[1 +2\cos\theta\,x_3 -\sin^2\theta\big(\cos^2\phi\,T_1\nonumber\\ &&\quad+\sin^2\phi\,T_2\big) -\cos^2\theta\,T_3\big],\\
    R_{10}(\theta,\phi)&=&\frac{1}{4}\big[1 -2\cos\theta\,x_3 -\sin^2\theta\big(\cos^2\phi\,T_1\nonumber\\ &&\quad+\sin^2\phi\,T_2\big) -\cos^2\theta\,T_3\big].\nonumber
\end{eqnarray}
The corresponding coefficients $\qty{R_{i_A}(\theta,\phi),R_{j_B}(\theta,\phi)}$ for the reduced matrices are given by
\begin{equation}
\begin{split}
    R_{0_A}(\theta)&=\frac{1}{2}\qty(1+x_3\cos\theta) =R_{1_B}(\theta),\\
    R_{1_A}(\theta)&=\frac{1}{2}\qty(1-x_3\cos\theta)=R_{0_B}(\theta).
\end{split}
\end{equation}

The necessary minimization to determine the optimal computational basis leads again to the same three cases that we had for the symmetric case \eqref{eq:Cases-Angles-Xs}. By defining the function
\begin{align}\label{eq:g-}
     g_{-}\equiv\;& \frac{1 - T_3 + 2 x_3}{4}\log_2\qty[\frac{1 - T_3 + 2 x_3}{(1+x_3)^2}]\nonumber\\
    &\;\;+
    \frac{1 - T_3 - 2 x_3}{4}\log_2\qty[\frac{1 - T_3 - 2 x_3}{(1-x_3)^2}]\nonumber\\ 
    &\;\;+ \frac{1-T_3}{2}\log_2\qty(\frac{1-T_3}{1-x_3^2}),
\end{align}
the classical correlations' quantifier \eqref{eq:CorrClasicas} is given by
\begin{equation}\label{eq:CorrClas-XAS}
\begin{split}
    \mathcal{C}\qty(\rho^{X_{as}}) &= \min\qty(g_1,g_2,g_{-}),
\end{split}
\end{equation}
with $g_1$ and $g_2$ already defined in \eqref{eq:g1(T1)} and \eqref{eq:g2(T2)}, respectively. As before, the function corresponding to the minimization defines the angles $\theta$ and $\phi$ for the optimal computational basis:
\begin{equation}\label{eq:OptimalBasis-Angles-Xas}
    \begin{cases}
    &\mathcal{C}\qty(\rho^{X_{as}}) = g_{-}\;\Longrightarrow\; \theta=0\qc\phi\in\qty[0,2\pi],\\
    &\mathcal{C}\qty(\rho^{X_{as}}) = g_{1}\;\Longrightarrow\;\theta=\frac{\pi}{2}\qc\phi=0,\\
    &\mathcal{C}\qty(\rho^{X_{as}}) = g_{2}\;\Longrightarrow\;\theta=\frac{\pi}{2}\qc\phi=\frac{\pi}{2}.
    \end{cases}
\end{equation}
As we did before, we rewrite the state $\rho^{X_{as}}$ in the optimal computational basis with the determined $\theta$ and $\phi$ angles and determine the corresponding probability distributions $P^{\qty(\theta,\phi)}\qty(i_A,j_B,\Phi)$ \eqref{eq:P_phi} as well as the marginal distributions $P_A^{\qty(\theta,\phi)}\qty(i_A,\Phi)$ and $P_B^{\qty(\theta,\phi)}\qty(i_B,\Phi)$. The maximization procedure for these probability distributions is analogous to the one performed for the symmetric case and it is straightforward to realize that the LAQC quantifier for anti-symmetric X states is given by
\begin{equation}\label{eq:LAQC-EstadosX-as}
    \mathcal{L}\qty(\rho^{X_{as}}) = \max\qty(g_1,g_2,g_{-}).
\end{equation}

\subsubsection{A single-parameter anti-symmetric X state}

In \cite{Verstraete-MAF}, Verstraete et al. introduced a single parameter mixed 2-qubit state, defined as:
\begin{equation}\label{eq:Ejemplo-X_AS}
\rho_v = F \dyad{\Phi^+}+(1-F)\dyad{01},
\end{equation}
\noindent{}where $\ket{\Phi^+}=\frac{1}{\sqrt{2}}\qty(\ket{00}+\ket{11})$ and $F\in[0,1]$. It is straightforward to verify that this state is an anti-symmetric X state, with Bloch parameters
\begin{equation}\label{eq:Bloch-Ejemplo-X_AS}
\begin{split}
    x_3 = -y_3= 1-F\,\qc &T_1=-T_2=F,\\ 
    &T_3=2F-1,   
\end{split}
\end{equation}
and all others equal to zero. By substituting these parameters in \eqref{eq:g1(T1)} and \eqref{eq:g-}, we have that
\begin{subequations}\label{eq:g-g1-EjXas}
\begin{align}
    g_{-}^{v}(F)=&\,2-F -F\log_2\qty(2-F)\nonumber\\
    &\;+(1-F)\log_2\qty[\frac{1-F}{(2-F)^2}],\label{eq:g-_EjXas}\\
    g_{1}^{v}(F)=&,\frac{1}{2}\big[(1-F)\log_2(1-F)\nonumber\\
    &\qq{} +(1+F)\log_2(1+F)\big].\label{eq:g1-EjXas}
\end{align}
\end{subequations}
These are the same functions \eqref{eq:g+-EjXS} and \eqref{eq:g1-EjXS} that we obtained for state \eqref{eq:Ejemplo-X_S}. Therefore, we have that
\begin{equation}\label{eq:C&L-Ejemplo-X_AS}
    \mathcal{C}\qty(\rho_{v}) = g_{-}^{v}(F)\qc \mathcal{L}\qty(\rho_{v}) = g_{1}^{v}(F).
\end{equation}

The Concurrence for state \eqref{eq:Ejemplo-X_AS} is again given by \eqref{eq:Concurrencia_Ejemplo-X_S} and the QD is the same as in the one-parameter symmetric X state presented in \eqref{eq:Ejemplo-X_S}. Therefore, the corresponding graph is the same as the one in Figure \ref{fig:Ejemplo-X_S}.

\section{Conclusions}
We have studied the local available quantum correlations (LAQC) \cite{LAQC} of the subset of 2-qubit X states, where their local Bloch vectors have equal magnitude. Such states can be regarded as symmetric or anti-symmetric X states whether their local Bloch vectors are parallel or anti-parallel, respectively. We obtained exact analytical expressions for their classical correlations and LAQC quantifiers in terms of the Bloch parameters of the state. As examples of symmetric X states, we analyzed the (Markovian) amplitude damping decoherence of Werner states and found that their LAQC does not suffer sudden death under this quantum channel. Also, we studied families of one-parameter X states, both symmetric and anti-symmetric. For such states, there is less LAQC than entanglement (as measured via concurrence) as well as quantum discord.

\section*{Acknowledgments}
The authors are thankful to Gloria Buendia for her comments and discussions. Albrecht and Bellorín would also like to thank the support given by the research group GID-30 \emph{Teoría de Campos y Óptica Cuántica} at the Universidad Simón Bolívar, Venezuela. This work was partially funded by the \emph{2020 BrainGain Venezuela} grant awarded to H. Albrecht by the \emph{Physics without Frontiers} program of the ICTP. 

\appendix*
\section{Quantum Discord of 2-qubit symmetric and anti-symmetric X states}

In \cite{Li-QD-Xstates}, Li et al. presented an analytic expression for the quantum discord of X states with parallel local Bloch vectors. Their analysis allows to compute the classical correlations $\mathcal{C}_D\qty(\rho^X)$ when you rewrite \eqref{eq:QD-Def} as
\begin{equation}
    D_A\qty(\rho^X)= I\qty(\rho^X) -\mathcal{C}_D\qty(\rho^X),
\end{equation}
where
\begin{equation}
    \mathcal{C}_D\qty(\rho^X)\equiv\min_{\qty{\Pi_i^A}}I[(\Pi^A\otimes\mathbbm{1}_2)\rho_{AB}].
\end{equation}

Lets start by defining monotonically decreasing function $f$ for $x\in[0,1]$:
\begin{equation}
    f(x)= -\frac{1+x}{2}\log_2(1+x) -\frac{1-x}{2}\log_2(1-x),
\end{equation}
so that the von Neumann entropy of the reduced matrices of $\rho^X$ is given by
\begin{equation}
S\qty(\rho_A^X) = 1 +f\qty(x_3)\qc
S\qty(\rho_B^X) = 1 +f\qty(y_3).
\end{equation}

Since the eigenvalues of $\rho^X$ \eqref{eq:estadosX-Bloch} are given
\begin{equation}
\begin{aligned}
\lambda_{1,2} &=\frac{1}{4}\qty[1-T_3\pm\sqrt{\qty(x_3-y_3)^2+\qty(T_1+T_2)^2}],\\
\lambda_{3,4}&=\frac{1}{4}\qty[1+T_3\pm\sqrt{\qty(x_3+y_3)^2+\qty(T_1-T_2)^2}],
\end{aligned}
\end{equation}
the mutual information \eqref{eq:InfoMutua} is then
\begin{equation}
    I\qty(\rho^X) = 2+f(x_3)+f(y_3) +\sum_{i=1}^{4}\lambda_i\log_2\lambda_i.
\end{equation}
As for $\mathcal{C}_D\qty(\rho^X)$, it is given by
\begin{equation}
    \mathcal{C}_D\qty(\rho^X) = 1-f(x_3) -\min\qty{S_1,S_2,S_3},
\end{equation}
where
\begin{equation}
S_1 \equiv -\sum_{i=1}^{4}\rho^X_{ii}\log_2\qty[\frac{2\rho_{ii}^X}{1-(-1)^{i}y_3}]
\end{equation}
and
\begin{equation}
\begin{aligned}
\rho^X_{\smqty{11\\22}}=\frac{1}{4}(1+x_3\pm{}y_3\pm{}T_3),
\\
\rho^X_{\smqty{33\\44}}=\frac{1}{4}(1-x_3\pm{}y_3\mp{}T_3),
\end{aligned}
\end{equation}
and
\begin{equation}
    S_{2,3}\equiv 1+f\qty(\sqrt{x_3^2+T_{1,2}^2}).
\end{equation}

For X states where $\abs{x_3}=\abs{y_3}$, we have that $f(x_3)=f(y_3)$ so that
\begin{equation}
    I\qty(\rho^{X_s}) = 2+2f(x_3) +\sum_{i=1}^{4}\lambda_i\log_2\lambda_i,
\end{equation}
where
\begin{equation}
\begin{aligned}
\lambda_{1,2}^s &=\frac{1}{4}\qty[1-T_3\pm\qty(T_1+T_2)],\\
\lambda_{3,4}^s&=\frac{1}{4}\qty[1+T_3\pm\sqrt{4x_3^2+\qty(T_1-T_2)^2}],
\end{aligned}
\end{equation}
for $x_3=y_3$ (symmetric) and
\begin{equation}
\begin{aligned}
\lambda^{as}_{1,2} &=\frac{1}{4}\qty[1-T_3\pm\sqrt{4x_3^2+\qty(T_1+T_2)^2}],\\
\lambda^{as}_{3,4}&=\frac{1}{4}\qty[1+T_3\pm\qty(T_1-T_2)],
\end{aligned}
\end{equation}
for $x_3=-y_3$ (anti-symmetric). As for $S_1$, we have that for $\rho^{X_s}$
\begin{equation}
\begin{aligned}
\rho^{X^s}_{11}&=\frac{1}{4}(1+2x_3+T_3)\qc
\rho^{X^s}_{22}=\frac{1}{4}(1-T_3),
\\
\rho^{X^s}_{33}&=\frac{1}{4}(1-T_3)\qc
\rho^{X^s}_{44}=\frac{1}{4}(1-2x_3+T_3),
\end{aligned}
\end{equation}
and
\begin{equation}
\begin{aligned}
\rho^X_{11}&=\frac{1}{4}(1+T_3)\qc
\rho^X_{22}=\frac{1}{4}(1+2x_3-T_3),
\\
\rho^X_{33}&=\frac{1}{4}(1-2x_3-T_3)\qc
\rho^X_{44}=\frac{1}{4}(1+T_3),
\end{aligned}
\end{equation}
for $\rho^{X_{as}}$.

As for the examples presented in this paper, it is straightforward to verify that for all three cases \eqref{eq:WernerAD-Bloch}, \eqref{eq:Ejemplo-X_S}, and \eqref{eq:Ejemplo-X_AS},
\begin{equation}
    \min\qty{S_1,S_2,S_3}=S_2=S_3,
\end{equation}
so that
\begin{equation}\label{eq:QD-Examples}
\begin{aligned}
    D_A\qty(\rho_{AB}) &=3+f(x_3)+f\qty(\sqrt{x_3^2+T_2^2})\\
    &\;\;+\sum_{i=1}^{4}\lambda_i\log_2\lambda_i,
\end{aligned}
\end{equation}
where the Bloch parameters were given in \eqref{eq:WernerAD-Bloch}, \eqref{eq:Bloch-Ejemplo-X_S}, and \eqref{eq:Bloch-Ejemplo-X_AS}, respectively.

\bibliographystyle{unsrt}
\bibliography{biblio-qit}
\include{biblio-qit}

\end{document}